# Study of polycrystalline bulk $Sr_3OsO_6$ double-perovskite insulator: comparison with 1000 K ferromagnetic epitaxial films

*Jie Chen,[1,2,*] Hai L. Feng,[3] Yoshitaka Matsushita,[4] Alexei A. Belik,[1] Yoshihiro Tsujimoto,[1,2] Masahiko Tanaka,[5] Duck Young Chung,[6] Kazunari Yamaura [1,2]*

[1] Research Center for Functional Materials, National Institute for Materials Science, 1-1 Namiki, Tsukuba, Ibaraki 305-0044, Japan

[2] Graduate School of Chemical Sciences and Engineering, Hokkaido University, North 10 West 8, Kita-ku, Sapporo, Hokkaido 060-0810, Japan

[3] Institute of Physics, Chinese Academy of Sciences, Beijing 100190, China

[4] Materials Analysis Station, National Institute for Materials Science, 1-2-1 Sengen, Tsukuba, Ibaraki 305-0047, Japan

[5] Synchrotron X-ray Station at SPring-8, National Institute for Materials Science, Kouto 1-1-1, Sayo-cho, Hyogo 679-5148, Japan

[6] Materials Science Division, Argonne National Laboratory, Lemont, Illinois 60439, United States

**Abstract**

Polycrystalline $Sr_3OsO_6$, which is an ordered double-perovskite insulator, is synthesized via solid-state reaction under high-temperature and high-pressure conditions of 1200 °C and 6 GPa. The synthesis enables us to conduct a comparative study of the bulk form of $Sr_3OsO_6$ toward revealing the driving mechanism of 1000 K ferromagnetism, which has recently been discovered for epitaxially grown $Sr_3OsO_6$ films. Unlike the film, the bulk is dominated by antiferromagnetism rather than ferromagnetism. Therefore, robust ferromagnetic order appears only when $Sr_3OsO_6$ is under the influence of interfaces. A specific heat capacity of $39.6(9) \times 10^{-3}$ J $mol^{-1}$ $K^{-2}$ is found at low temperatures (<17 K). This value is remarkably high, suggesting the presence of possible fermionic-like excitations at the magnetic ground state. Although the bulk and film forms of $Sr_3OsO_6$ share the same lattice basis and electrically insulating state, the magnetism is entirely different between them.

# 1. Introduction

New materials research and the functional development of perovskite-related 5d oxides are ongoing because of the growing prospects for new and potentially useful characteristics. For example, a high magnetic coercivity up to 55 T in $Sr_3NiIrO_6$,[1] the bulk exchange bias of $Ba_2FeOsO_6$,[2] the Slater/Lifshitz transition of $NaOsO_3$ [3-7] and $Cd_2Os_2O_7$,[8, 9] the Anderson and Blount-type transition of $LiOsO_3$,[10, 11] and the enhanced spin-phonon-electronic coupling of $NaOsO_3$ [12] have all been discovered in the past decade. These characteristics may share a fundamental background including large 5d orbitals, strong spin-orbit coupling (SOC), and the non-trivial relativistic effect.[13]

Most recently, high-temperature ferromagnetism above 1000 K has been observed for an epitaxially grown film of the double-perovskite oxide $Sr_3OsO_6$ (= $Sr_2Sr^{2+}Os^{6+}O_6$), in which $Os^{6+}$ and $Sr^{2+}$ atoms are ordered at the perovskite B-site.[14] The film is highly electrically insulating; therefore, the ferromagnetic (FM) transition temperature is the highest among oxides and insulators.[14] The magnetic ground state comprises a canted FM order of Os-moments (~0.77 $\mu_B$/Os).[14] The remarkable magnetic property is very promising for spintronics applications and beyond.

In solid-state chemistry, the double-perovskite oxide $Sr_3OsO_6$ can be categorized into a class termed "single-magnetic sublattice", like $Sr_2MOsO_6$ ($M$ = Y, In, Sc, Li, Na, Mg, Ca).[15] Although double-magnetic sublattice members like $Sr_2MOsO_6$ ($M$ = Cr, Fe, Co, Ni, Cu) display varied magnetic order depending on the combination of magnetic elements,[16-20] single-magnetic sublattice members display only antiferromagnetic (AFM) order at low temperatures.[15] The highest magnetic transition temperature to date is at most 110 K for $Sr_2MgOsO_6$; therefore, the 1000 K FM order of $Sr_3OsO_6$ is highly unprecedented because it is clearly beyond the present scheme of the single-magnetic sublattice class. Currently, the driving mechanism of the robust FM order is unclear, and a $Sr_3OsO_6$ bulk of a polycrystal or single crystal has not been achieved. Synthesis of the $Sr_3OsO_6$ bulk is imperative to investigate this mechanism.

In this study, we successfully synthesize polycrystalline $Sr_3OsO_6$ via a high-pressure method. The electrical and structural properties are comparable with those of the epitaxial film. Surprisingly, the robust long-range FM order is not established at temperatures above 2 K in the $Sr_3OsO_6$ bulk. Herein, we report the structural and magnetic properties of the $Sr_3OsO_6$ bulk synthesized for the first time.

## 2. Experimental

Polycrystalline $Sr_3OsO_6$ was synthesized under a high-pressure condition by solid-state reaction from powders of $SrO_2$ (laboratory-made from 99.0% $SrCl_2·6H_2O$, Wako Pure Chem. Co.),[21] SrO (laboratory-made from 99.9% $SrCO_3$, Wako Pure Chem. Co.), and $OsO_2$ (laboratory-made from 99.95% Os, Nanjing Dongrui Platinum Co., Ltd.). The powders were thoroughly mixed at the stoichiometric ratio $2SrO + SrO_2 + OsO_2$, and then sealed in a Pt capsule. The procedure was conducted in an Ar-filled glove box. Under a pressure of 6 GPa, using a multi-anvil-type apparatus (CTF-MA1500P, C&T Factory Co., Ltd, Japan), the capsule was heated at 1200 °C for 30 min, after which the pressure was gradually released. The obtained pellet was dense and black in color.

The high-pressure-synthesized $Sr_3OsO_6$ was investigated by X-ray diffraction (XRD) using Cu-Kα radiation with a commercial apparatus, RIGAKU-MiniFlex 600. The XRD measurement was conducted using a fresh product synthesized on the same day because we found that the product tended to gradually decompose even at room temperature when it was finely ground. Although the chemical nature was not fully revealed, we conducted synchrotron XRD measurements at the X-ray Science Division beamline in the Advanced Photon Source, USA. Unfortunately, we failed to obtain a synchrotron XRD pattern of sufficiently high quality, most likely due to the chemical instability. We later conducted the synchrotron XRD measurements again using the high-precision powder X-ray diffractometer, which is installed at the BL15XU beamline, SPring-8, Japan.[22, 23] For the measurements, an as-made product was taken out from the Pt capsule in an Ar atmosphere, finely crushed, and then sealed into a "Lindemann" quartz capillary without exposure to air. A Rietveld analysis using the RIETAN-FP and VESTA software [24, 25] was conducted on the synchrotron XRD pattern.

The physical properties were measured as detailed below using several pieces cut from the polycrystalline $Sr_3OsO_6$; we did not use the ground powder. The electrical resistivity ($\rho$) of polycrystalline $Sr_3OsO_6$ was measured at temperatures between 150 K and 400 K by the four-probe method in a physical property measurement system (PPMS, Quantum Design, Inc.). The gauge current was 1 mA. Below 150 K, $\rho$ was too high to be properly measured in the system. The specific heat capacity ($C_p$) was measured in the PPMS by a thermal relaxation method from 300 K to 2 K with Apiezon N grease, which thermally connects the material to the holder stage. The $C_p$ measurement was conducted with and without applying a magnetic field of 10 kOe.

The magnetic properties were measured in a magnetic property measurement system (MPMS, Quantum Design, Inc.). The magnetic susceptibility ($\chi$) was measured at temperatures between 2 K and 400 K in an applied magnetic field ($H$) of 10 kOe under field-cooling (FC) and zero-field-cooling (ZFC) conditions. The complex magnetic susceptibility ($\chi_{ac} = \chi' + i\chi''$) was measured between 2 K and 50 K at a variety of frequencies ($f$) in an ac magnetic field ($H_{ac}$) of 5 Oe without an applied static field (except the geomagnetic field). The isothermal magnetization ($M$) was measured for a sweeping field from -70 to +70 kOe at 5 K.

## 3. Results and discussion

The XRD pattern (Fig. 1) confirms the quality of the high-pressure product. The lattice parameters of a temporarily assumed tetragonal cell are $a$ = 5.827(16) Å and $c$ = 8.275(6) Å. The parameters $\sqrt{2}a$ [= 8.24(2) Å] and $c$ are comparable to the spacing distances of out-of-plane 8.22(3) Å and in-plane 8.24(3) Å of the $Sr_3OsO_6$ film, respectively.[14] The peak at $2\theta \approx$ 18-19° with *hkl* indices *101* suggests a high degree of order of the Sr and Os atoms at the perovskite B site.[26] As the Sr and Os order is confirmed to be as high as that of the film,[14] we conclude that the film and bulk $Sr_3OsO_6$ share the same lattice basis.

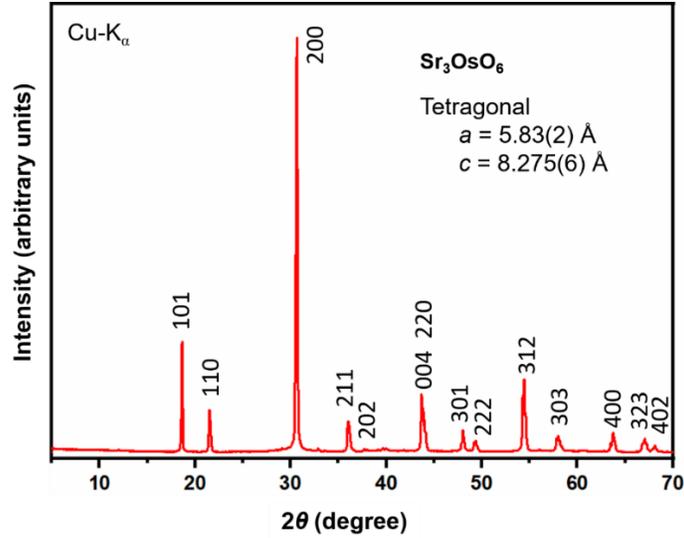

Fig. 1  Powder XRD pattern for $Sr_3OsO_6$ synthesized under high pressure. A tetragonal cell is temporarily applied for comparison. The measurement was conducted on the same day as the synthesis of the compound.

To investigate the structure further, Rietveld analysis is conducted on the synchrotron XRD pattern (Fig. 2 and Table 1). In the initial analysis, several possible structural models with different space groups such as *I*4/*m* (No. 87), *I*4$_1$/*a* (No. 88), *P*2$_1$/*n* (No. 14), *C*2/*m* (No. 12), and *P*-1 (No. 2) are considered.[20, 27-30] When we assume the tetragonal models (*I*4/*m* and *I*4$_1$/*a*), the Bragg reflections do not exactly match with the observed peaks. Therefore, we consider models with a lower symmetry than the tetragonal. Note that the structural symmetry of the $Sr_3OsO_6$ film was not exactly identified probably because of technical difficulty in analyzing the epitaxial film.[14]

The monoclinic models (*P*2$_1$/*n* and *C*2/*m*) are tested as well. An improvement of the refinement is achieved to some extent; however, few small peaks still do not exactly match with those in the models. This issue is overcome by employing the triclinic model (*P*-1), which was proposed for $Ba_2LaRuO_6$.[30] Although there is no dramatic improvement in the refinement, the *P*-1 model is the best to describe the structure of the $Sr_3OsO_6$ bulk, as argued previously for $Ba_2LaRuO_6$.[30]

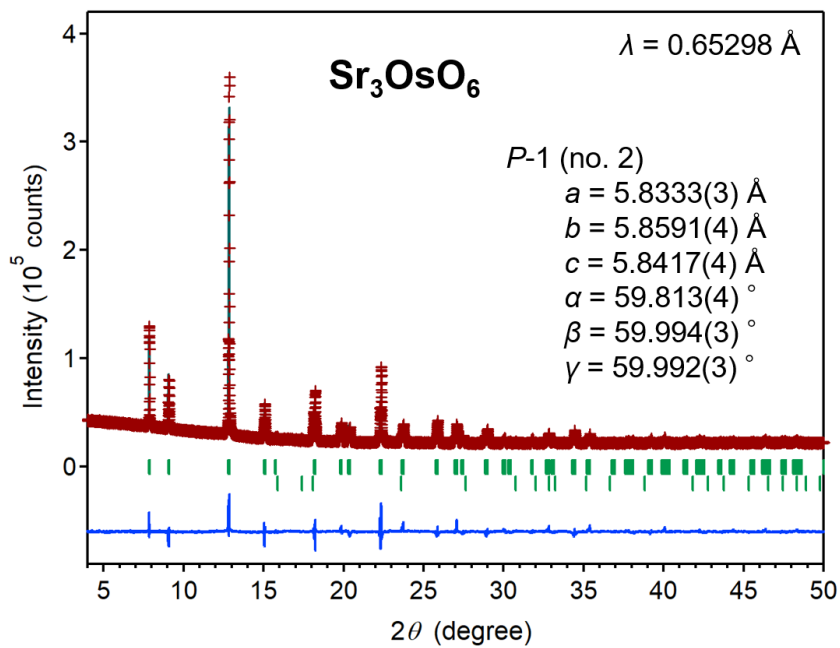

Fig. 2   Rietveld refinement of the synchrotron XRD profile ($\lambda = 0.65298$ Å) collected at room temperature for a high-pressure synthesized $Sr_3OsO_6$. A triclinic ($P$-1) model and an Os impurity were considered in the analysis. The crosses and solid lines show the observed and calculated patterns, respectively, with their differences shown at the bottom. The expected Bragg reflections are marked by ticks for the $P$-1 structure (top; 98.48 wt.%) and Os (bottom; 1.52 wt.%).

In the final refinement, the oxygen coordination was free at the beginning; however, it was challenging to obtain a reasonable solution. For example, the calculated bond lengths of Sr–O and Os–O were too unusual to accept. To ameliorate the situation, it was necessary to fix the oxygen coordination at that of $Ba_2LaRuO_6$.[30] Besides, refinement of the isotropic atom displacement parameters ($U_{eq}$) of the Sr, Os, and O atoms was also very challenging. We had to apply a constraint to the parameters, like $U_{eq}(Sr1) = U_{eq}(Sr2) = U_{eq}(Os1)$. Without this constraint, the parameters would never have been acceptable. This indicates that the analysis is affected by the quality of the synchrotron XRD pattern.

Table 1  Atomic coordinates and equivalent isotropic displacement parameters ($U_{eq}$, $10^{-3}$ Å$^2$) for a triclinic Sr$_3$OsO$_6$ as measured by the synchrotron X-ray diffraction at room temperature

| Site | WP | Occp. | $x$ | $y$ | $z$ | $U_{eq}$ |
|---|---|---|---|---|---|---|
| Sr1 | 2$i$ | 1 | 0.295(2) | 0.1759(9) | 0.2834(19) | 8.31(17) |
| Sr2 | 1$a$ | 1 | 0 | 0 | 0 | 8.31 |
| Os  | 1$h$ | 1 | 0.5 | 0.5 | 0.5 | 8.31 |
| O1  | 2$i$ | 1 | 0.7166 | 0.3254 | 0.2246 | 7.09 |
| O2  | 2$i$ | 1 | 0.2980 | 0.2450 | 0.7040 | 7.09 |
| O3  | 2$i$ | 1 | 0.2372 | 0.7478 | 0.3092 | 7.09 |

WP: Wyckoff position. Space group: $P$-1 (Triclinic; No.2); lattice constants $a = 5.8333(3)$ Å, $b = 5.8591(4)$ Å, $c = 5.8417(4)$ Å, $\alpha = 59.813(4)$ °, $\beta = 59.994(3)$ °, and $\gamma = 59.992(3)$ °; cell volume = 140.96(2) Å$^3$; $d_{cal} = 6.468$ g cm$^{-3}$; Chemical formula sum: Sr$_3$OsO$_6$ ($Z = 1$); and the final $R$ values are 3.776% ($R_{wp}$), 2.041% ($R_p$), 4.031% ($R_B$), and 1.534% ($R_F$). The coordination ($x$, $y$, $z$) and temperature factor of the oxygen were fixed at the values of Ba$_2$LaRuO$_6$.[30]

The occupancy factors of metal and oxygen were fixed at full in the refinement because the synthesis did not suggest any presence of significant non-stoichiometry. Indeed, a preliminary analysis with unfixed occupancies did not help to improve the refinement. The structure images are drawn using the refined parameters in Figs. 3a and 3b, and the selected bond lengths in the triclinic structure are summarized in Table S1.

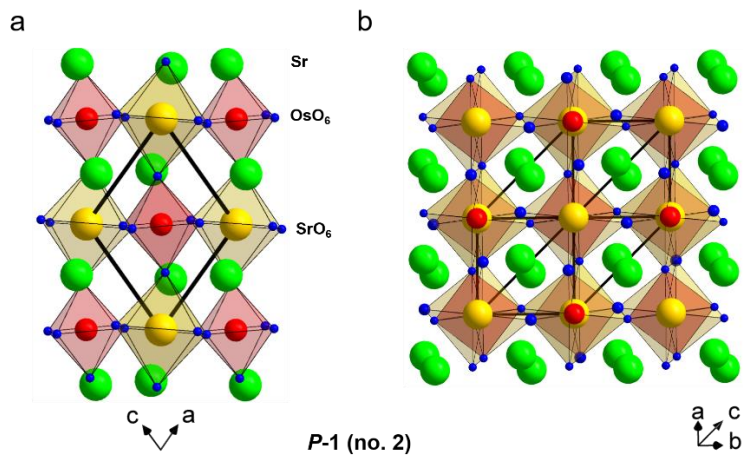

Fig. 3    (a) A crystal structure image of the triclinic $Sr_3OsO_6$ (*P*-1) viewed from the b-axis, in which $SrO_6$ and $OsO_6$ octahedra are drawn in orange and red, respectively. The Sr atom position at the perovskite A-site is drawn as a green sphere. (b) Alterative image viewed from the [1 1 -1] direction.

In addition, we analyzed the synchrotron XRD pattern by considering a cubic structure coexisting with a tetragonal structure in the high-pressure product; surprisingly, an improvement in refinement quality was achieved. This suggests $Sr_3OsO_6$ crystalizes into multiple symmetry structures under the high-pressure condition (see Fig. S1 and Tables S2 and S3); otherwise, a metastable structure possibly appeared during the measurement by a partial decomposition. The structure image each was drawn using the refined parameters (Figs. S2 and S3) for a reference. Based on the structural analysis, we concluded that a conclusive analysis should be conducted using a single crystal rather than powder.

Although no significant discrepancy was detected between the bulk and film structures of $Sr_3OsO_6$, a remarkable difference was observed in the magnetic properties. Figure 4a shows the temperature dependence of $\chi$ in the $Sr_3OsO_6$ bulk; it does not indicate the establishment of a long-range FM order at temperatures above 2 K. This strictly contrasts what was observed for the $Sr_3OsO_6$ film.[14] Moreover, a small anomaly is detected at ~12 K with thermal hysteresis, suggesting a glassy transition or FM transition. It should be noted that $Sr_5Os_3O_{13}$,[31] $Sr_2Os_3O_5$,[32] and $Sr_7Os_4O_{19}$ [32] have been synthesized in the ternary system Sr–Os–O, but none of them displays a magnetic anomaly at ~12 K.

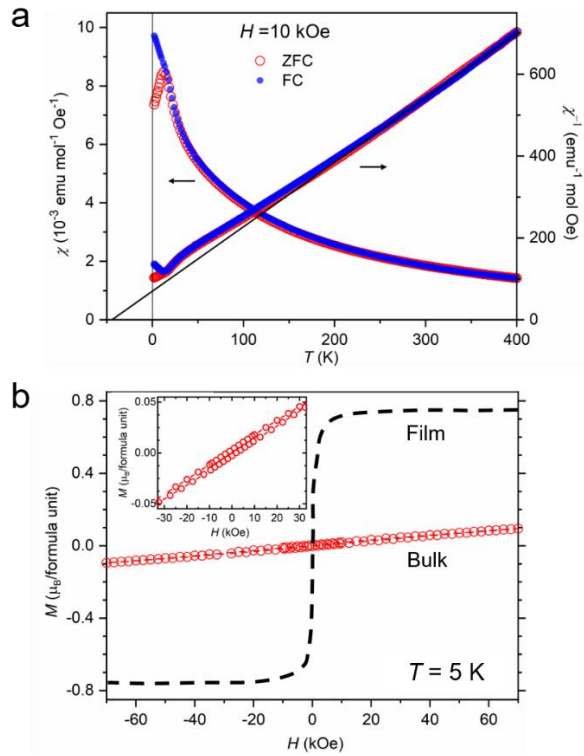

Fig. 4 (a) Temperature dependence of $\chi$ of polycrystalline $Sr_3OsO_6$, measured in a field of 10 kOe. The solid line is simulated by the Curie-Weiss law. (b) Isothermal magnetizations at 5 K for the bulk (red markers) and film (broken curve; estimated from the data in Ref. 14) of $Sr_3OsO_6$. Inset is an expanded view of the bulk data.

To conduct a quantitative analysis, the Curie-Weiss law $1/\chi = (T-\theta_W)/C$ was applied to fit to the curve above ~200 K, where $\theta_W$ is the Weiss temperature and $C$ is the Curie constant. The estimated $\theta_W$ and $C$ are –47.0(5) K and 0.638(2) emu mol$^{-1}$ K, respectively. There is essentially no difference in the sets of parameters for the ZFC and FC curves. The $C$ value corresponds to the effective magnetic moment ($\mu_{eff}$) of 2.259(3) $\mu_B$, which is slightly higher than that of other $Os^{6+}$ oxides, such as 1.99 $\mu_B$ for $Ca_3OsO_6$ [27] and 1.867(4) $\mu_B$ for $Sr_2MgOsO_6$.[15] The slight overestimation is possibly caused by the impact of the SOC, as argued elsewhere.[33] Although the negative $\theta_W$ is much reduced from -151 K for $Ca_3OsO_6$ [27] and -347.3(5) K for $Sr_2MgOsO_6$,[15] the dominant interaction is still AFM.

The reduced $\theta_W$ indicates the presence of a weakened AFM interaction, for which the elongated distances between the nearest neighbor magnetic atoms of $Os^{6+}$ caused by the larger $Sr^{2+}$ rather than $Ca^{2+}/Mg^{2+}$ are likely responsible. Besides, we studied the relation among the distance between the nearest neighbor Os atoms and the magnetic transition temperature and $\theta_W$ of the double perovskite Os oxides having a single-magnetic sublattice; however, any comprehensive trend could not be established, unfortunately (see Table 2).

Table 2  Comparison of distances between the neighbor Os atoms ($d_{Os-Os}$) and magnetic transition temperatures ($T_{mag}$) of double perovskite Os oxides having a single-magnetic sublattice

|  | Compound | S.G. | $T_{mag}$ (K) | $d_{Os-Os}$ (Å) | $\theta_W$ (K) | Refs. |
|---|---|---|---|---|---|---|
| $Os^{5+}$ | $Ca_2InOs^{5+}O_6$ | $P2_1/n$ | 14 | 5.49–5.68 | -77 | 34 |
|  | $La_2NaOs^{5+}O_6$ | $P2_1/n$ | 17 | 5.61–5.95 | -74 | 35 |
|  | $Ba_2YOs^{5+}O_6$ | $Fm$-$3m$ | 69 | 5.91 | -772 | 36 |
| $Os^{6+}$ | $Ca_2MgOs^{6+}O_6$ | $P2_1/n$ | 12 | 5.41–5.54 | -71.5(7) | 37 |
|  | $Sr_2MgOs^{6+}O_6$ | $I4/m$ | 110 | 5.56–5.58 | -347.3(5) | 37 |
|  | $Ca_3Os^{6+}O_6$ | $P2_1/n$ | 50 | 5.66 | -151 | 38 |
|  | $Sr_3Os^{6+}O_6$ | $P$-1 | 12 | 5.83–5.86 | -47.0(5) | this work |
|  | $Ba_2CaOs^{6+}O_6$ | $Fm$-$3m$ | 50 | 5.91 | -156.2(3) | 39, 40 |
| $Os^{7+}$ | $Ba_2LiOs^{7+}O_6$ | $Fm$-$3m$ | 8 | 5.73 | -40.48 | 41 |
|  | $Ba_2NaOs^{7+}O_6$ | $Fm$-$3m$ | 8 | 5.86 | -32.45 | 41 |

S.G.: Space group; $d_{Os-Os}$: Distance between the neighbor Os atoms; $T_{mag}$: magnetic transition temperature

To further confirm the presence of a possible long-range FM order, the isothermal magnetization was measured at 5 K (Fig. 4b). Although the magnetization reached ~0.1 $\mu_B$/Os at 5 K in an applied magnetic field of 70 kOe, it was less than one-seventh of that of the $Sr_3OsO_6$ film (~0.77 $\mu_B$/Os). Besides, a step-like behavior, which is commonly observed for an FM material, was not visible, contrasting what was observed for the $Sr_3OsO_6$ film (refer to the varying curves in Fig. 4b). The measurement again does not support the presence of a long-range FM order in the $Sr_3OsO_6$ bulk. A weak magnetic hysteresis (see the inset of Fig. 4b) likely links to a magnetic glassy transition at ~12 K, suggesting some degree of FM interaction may be present in the bulk compound. We investigate the glassy transition further in the next.

We studied the 12 K anomaly in the $\chi$-$T$ curves by means of a $\chi_{ac}$ measurement. Figures 5a and 5b show real and imaginary components of $\chi_{ac}$, respectively. A peak-like feature is clearly observed in both the $\chi'$-$T$ and $\chi''$-$T$ curves, suggesting it associates with the 12 K anomaly. The peak top position shows a frequency dependence; it shifts toward a higher temperature with increasing frequency. Because such $f$ dependency is a common phenomenon of magnetically glassy transition,[42-46] we can reasonably assume that a magnetically glassy state is established at ~12 K on cooling in the $Sr_3OsO_6$ bulk. Note that we are unable to deduce the volume fraction of the glassy state in the bulk from the $\chi_{ac}$ data.

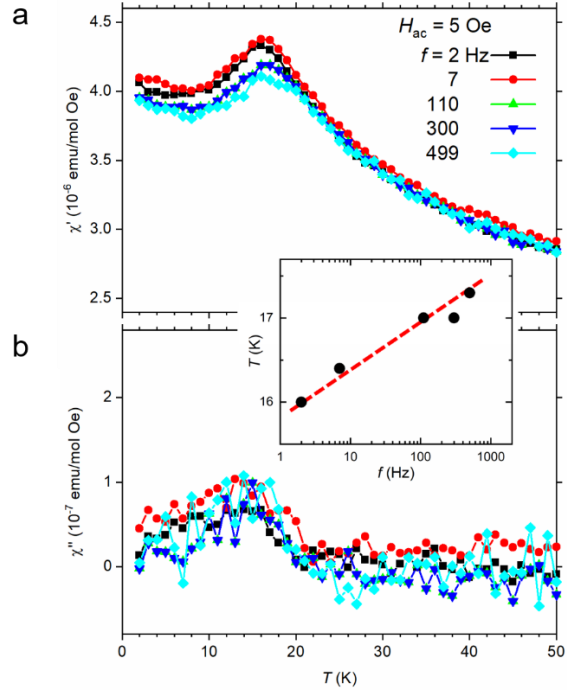

Fig. 5 (a) Temperature and frequency dependences of the real $\chi'$ and (b) imaginary $\chi''$ components of $\chi_{ac}$ of $Sr_3OsO_6$ measured at a $H_{ac}$ of 5 Oe. (Inset) Frequency dependence of the peak top position in the real component data. The red broken line is a visual guide.

The charge transport of the $Sr_3OsO_6$ bulk was investigated via measurements of the temperature dependence of $\rho$ as shown in Fig. 6. Although the room-temperature $\rho$ (~1.5 kΩ·cm) is approximately 20 times higher than that of the $Sr_3OsO_6$ film, the semiconducting-like temperature dependence is analogous to what was observed for the film.[14] The rigid shift in the $\rho$ vs. $T$ curve is possibly due to the presence of chemical instability and grain boundaries. A quantitative study using a single crystal is, therefore, necessary.

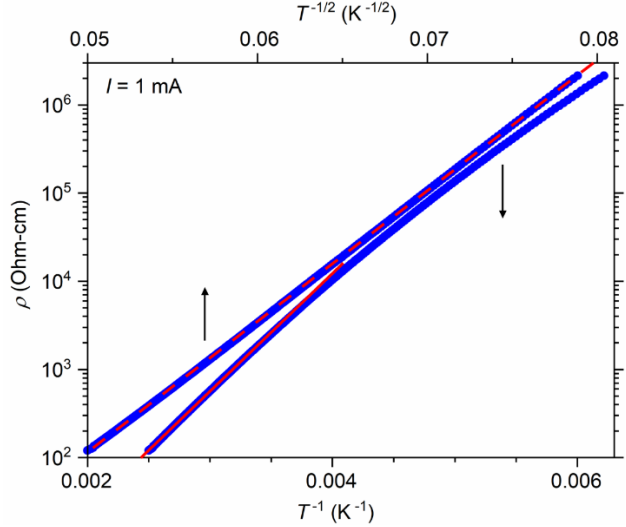

Fig. 6  Temperature dependence of $\rho$ of polycrystalline $Sr_3OsO_6$ upon cooling in two plot styles: $\rho$ vs. $T^{-1}$ and $\rho$ vs. $T^{-1/2}$. The corresponding plots upon heating do not show substantial differences (not shown). The red line indicates a fitting to the Arrhenius law in the high-temperature region. The broken line serves as a visual guide.

The present data are plotted in both the Arrhenius (logarithmic $\rho$ vs. $T^{-1}$, bottom axis) and the Efros-Shklovskii variable range hopping (VRH) forms (logarithmic $\rho$ vs. $T^{-1/2}$, top axis).[47, 48] They indicate that the Efros-Shklovskii conduction is much better to describe the observation over the temperature range between 160 K and 400 K, as the data curve is surprisingly linear (the red broken line serves as a visual guide). This suggests that conduction occurs possibly under the influence of disorder with long-range Coulomb interactions.[48] Note that we tested the conversional VRH model (logarithmic $\rho$ vs. $T^{-1/4}$) as well (not shown); however, fitting was poorer than that by the Efros-Shklovskii model.

Although the Arrhenius law $\rho = \rho_0 \exp(E_a/k_B T)$, in which $E_a$ denotes the activation energy and $\rho_0$ and $k_B$ are the temperature-independent constant and Boltzmann constant, respectively, is not the best to analyze the conduction over the entire temperature range, it can approximately estimate $E_a$ in the high-temperature range between 300 K and 400 K. Fitting to the curve as shown by the red solid line in Fig. 6 estimated the lower limit of $E_a$ as 266.3(3) meV.

We measured $C_p$ of the $Sr_3OsO_6$ bulk from 300 K to 2 K on cooling with and without applying a magnetic field of 10 kOe as shown in Fig. 7. There is no anomaly associated with a phase transition over the temperatures, indicating that the magnetic susceptibility anomaly at 12 K is unlikely caused by a long-range magnetic ordering. As already suggested, a freezing short-range magnetic order or an independent magnetic impurity may be responsible for the 12 K anomaly. A gap between the 2 sets of data measured with and without the magnetic field is not visible, supporting the pictures.

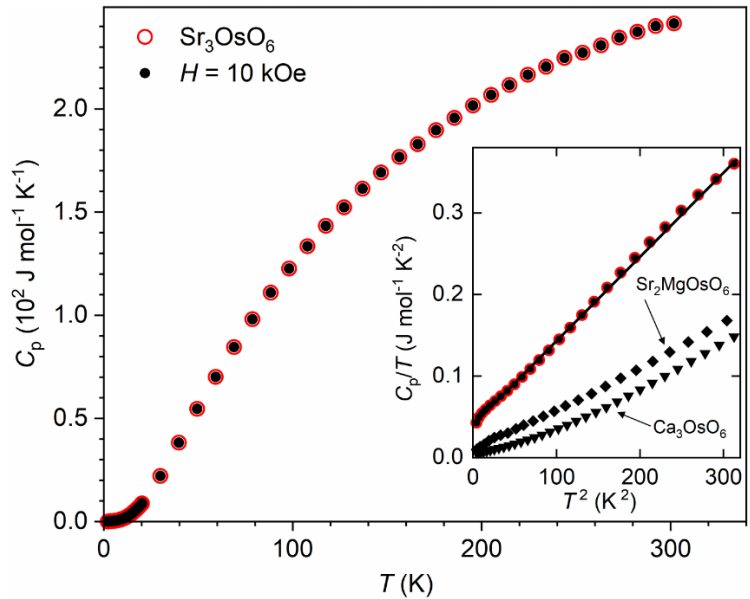

Fig. 7  $C_p$ vs. $T$ curve of $Sr_3OsO_6$ measured with and without applying a magnetic field of 10 kOe. The upper axis corresponds to the Dulong-Petit limit of the lattice specific heat. (Inset) A linear fit to the $C_p/T$ vs. $T^2$ curve near the low-temperature limit. The sets

of data for reference compounds (solid diamonds and triangles) are taken from Refs. [27] and 15, respectively.

Considering the low-temperature part (<17 K, inset to Fig. 7), the $C_p/T$ vs. $T^2$ plot displays a linear feature; therefore, we applied the approximated Debye model for quantitative analysis. The analytical formula $C_p/T = \gamma + \beta_0 T^2$, in which $\gamma$ and $\beta_0$ are the constants, is applied to the linear part, and $\gamma$ of 39.6(9)×10$^{-3}$ J mol$^{-1}$ K$^{-2}$ and $\beta_0$ of 10.3(6)×10$^{-4}$ J mol$^{-1}$ K$^{-4}$ are obtained. $\beta_0$ provides a Debye temperature of 266(1) K. The first-principles calculation revealed that the ground state is fully gapped [14] and the Sr$_3$OsO$_6$ bulk is indeed as electrically insulating as the film; therefore, the electronic contribution to $\gamma$ should be nearly zero because it is proportional to the charge density of state at the Fermi level. Even though some degree of hopping conduction is experimentally detected, the electronic contribution unlikely accounts for the large $\gamma$.

The experimental $\gamma$ is likely contributed by a magnetic origin regarding Os$^{6+}$ (5d$^2$). However, the $\gamma$ values of related 5d$^2$ insulators are not as prominent as that of Sr$_3$OsO$_6$. For example, the $\gamma$ values of Ca$_3$OsO$_6$ ($T_{AFM}$ = 50 K) [27] and Sr$_2$MgOsO$_6$ ($T_{AFM}$ = 110 K) [15] are 0.4×10$^{-3}$ and 0.1(5)×10$^{-3}$ J mol$^{-1}$ K$^{-2}$, respectively. Because no long-range AFM order is established in the Sr$_3$OsO$_6$ bulk (except the 12 K-anomaly with unknown volume fraction), the substantial $\gamma$ may indicate the occurrence of a fermionic-like excitation.[49-52] Similarly, a spin-liquid insulator displays a finite value of $\gamma$ with a linear trend in the $C_p/T$ vs. $T^2$ plot. For example, $\gamma$ values of 15×10$^{-3}$ and 168×10$^{-3}$ J mol$^{-1}$ K$^{-2}$ were observed for κ-(BEDT-TTF)$_2$Cu$_2$(CN)$_3$ [51, 52] and Ba$_3$NiSb$_2$O$_9$,[50] respectively.

Among the 5d oxide insulators, large $\gamma$ value has rarely been observed. The $S$ = 1/2 hyperkagome antiferromagnet Na$_4$Ir$_3$O$_8$, for example, shows a $\gamma$ of 2×10$^{-3}$ J mol$^{-1}$ K$^{-2}$,[53] which is not as large as that of Sr$_3$OsO$_6$. In addition, a magnetically glassy state has been argued to occur with a linear temperature dependence in the magnetic heat capacity.[43, 45, 46, 54] In this study, although glassy characteristics such as a broad cusp in the $C_p$ data are not distinct, further study is needed to determine if the glassy state contributes to the large $\gamma$.

## 4. Conclusion

Polycrystalline $Sr_3OsO_6$ shares the lattice basis with the epitaxially grown film. However, the 1000 K long-range FM order is not in the polycrystalline $Sr_3OsO_6$, as evidenced by, for example, the Os moment being only ~0.1 $\mu_B$ (in 70 kOe at 5 K) in stark contrast to the ~0.77 $\mu_B$ of the film. Besides, the Weiss temperature is -47.0(5) K, indicating that the dominant interaction is AFM. This result suggests that the 1000 K long-range FM order appears only when $Sr_3OsO_6$ is under the influence of interfaces. Further, we need to carefully evaluate the influence of SOC on the magnetic order, as it has been suggested to have a significant impact on the magnetic ground state of a $5d^2$ double perovskite.[33] Finally, while the driving mechanism of the 1000 K FM order remains to be clarified, the remarkably large $\gamma$ [= 39.6(9)×$10^{-3}$ J $mol^{-1}$ $K^{-2}$] poses an additional open question regarding the possibility of fermionic-like excitation, which has rarely been observed for 5d oxide insulators.

## ASSOCIATED CONTENT

**Supporting Information**

Tables showing additional crystallographic data obtained at room temperature from a polycrystal $Sr_3OsO_6$. Powder synchrotron XRD pattern analyzed by a coexisting tetragonal and cubic structures model for a polycrystal $Sr_3OsO_6$. This material is available free of charge online at http://pubs.acs.org.

## AUTHOR INFORMATION

**Corresponding Author**

* Jie Chen

Quantum Solid State Materials Group
Research Center for Functional Materials
National Institute for Materials Science

1-1-Namiki, Tsukuba, Ibaraki 305-0044, Japan

TEL: +81-29-851-3354 (Ext: 8660)

E-mail: is.jiechen@gmail.com

**Notes**

The authors declare no competing financial interests.

**Author Contributions**

The manuscript was written through contributions from all the authors. All authors have given their approval for the final version of the manuscript.

**Funding Sources**

This study was supported in part by JSPS KAKENHI Grant Number JP16H04501, a research grant from Nippon Sheet Glass Foundation for Materials Science and Engineering (#40-37), and Innovative Science and Technology Initiative for Security, ATLA, Japan.

**ACKNOWLEDGMENTS**

We thank Y. Ishii and H. Yoshida for the helpful discussion. The synchrotron radiation experiments were performed at the NIMS synchrotron X-ray station at SPring-8 with the approval of the Japan Synchrotron Radiation Research Institute (Proposal Numbers: 2019B4500, 2019A4501). Work at Argonne (XRD) is supported by the U.S. DOE, Office of Basic Energy Science, Materials Science and Engineering Division. Use of the Advanced Photon Source at Argonne National Laboratory was supported by the U.S. Department of Energy, Office of Science, Office of Basic Energy Sciences, under Contract No. DE-AC02-06CH11357.

**For Table of Contents Only**

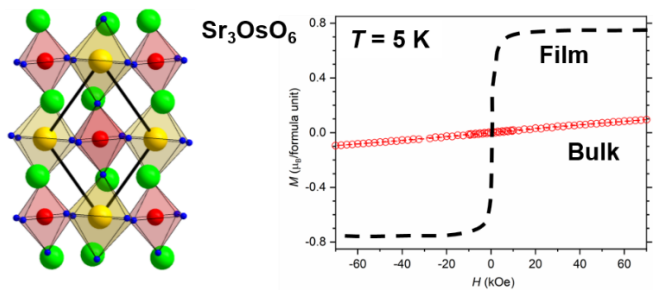

**TOC synopsis:**

$Sr_3OsO_6$ shows unprecedented ferromagnetism above 1000 K when it is in a film form, while it does not display major ferromagnetic character when it is in a polycrystal form. This mystery has not been solved, but polycrystalline $Sr_3OsO_6$ is revealed to be rather antiferromagnetic and shows the possibility of fermionic-like excitation at the magnetic ground state. The progress has been achieved by the successful synthesis of polycrystalline $Sr_3OsO_6$ under a high-pressure and high-temperature condition.